\documentclass[prd,eqsecnum,showpacs]{revtex4}
\usepackage{amsmath}
\usepackage{epsfig}

\begin{document}
\title{Renormalization Group approach to Inhomogeneous Cosmology }
\author{{J. Ib\'a\~nez and S. Jhingan}}
\affiliation{ Departamento de F\'{\i}sica Te\'orica, Universidad
del Pa\'{\i}s Vasco, Apartado 644, 48080, Bilbao, Spain }

\date{{\small \today}}

\begin{abstract}
Soliton solutions are recovered as scale-invariant asymptotic
states of vacuum inhomogeneous cosmologies using renormalization
group method. The stability analysis of these states is also given.
\end{abstract}

\pacs{98.80.Jk, 64.60.Ak}

\maketitle

\section{Introduction}

The study of influence of initial inhomogeneities upon the
evolution of cosmological models is an important issue in
Cosmology, both in order to understand the formation of
large-scale structures as well the smoothing away of these
inhomogeneities. One way to deal with the inhomogeneities is to
consider perturbations of known solutions, like
Friedmann-Robertson-Walker models or homogeneous Bianchi type
solutions. Another way is to assume inhomogeneous solutions from
the beginning and study their dynamical evolution; it is this that
we are concerned with here. The simplest inhomogeneous models are
the so called diagonal $G_2$ cosmologies for which the spacetime
admits two commuting Killing vectors whose orbits are
2-dimensional space-like surfaces. They represent inhomogeneities
of spatially homogeneous models and can be considered as
gravitational waves of a single polarization propagating over a
homogeneous background~\cite{A}. Using a few solution generating
techniques a large number of exact solutions with different
sources have been found~\cite{K}.

The asymptotic evolution of homogeneous Bianchi type models and
both their behavior near the initial singularity as well as their
future state have been widely studied~\cite{W}. One of the facts
relevant to our paper is that at late times Bianchi models can be
described as self-similar solutions. However, the asymptotic
behavior of inhomogeneous solutions, particularly $G_2$ metrics,
revealed a task much more difficult than that of homogeneous
models. Contrary to the homogeneous metrics for which the Einstein
equation reduce to an autonomous system of differential equations
that can be analyzed using techniques from the theory of dynamical
systems, the field equations of inhomogeneous spacetimes are
partial differential equations. In ref.~\cite{W} the evolution of
a particular class of $G_2$ metrics is studied showing that in
that case all the solutions are asymptotically self-similar. The
same result was obtained in~\cite{IO} and in~\cite{CC} where
special families of $G_2$ solutions were considered with a scalar
field. On the other side, it has been suggested that fluctuations
might evolve from arbitrary initial conditions to self-similar
form~\cite{C-C1989}. From all these results it is reasonable to
regard the self-similar solutions as describing the long time
asymptotic of inhomogeneous metrics. Hence it would be worthy to
analyze the asymptotic behavior of general class of $G_2$ metrics
by using a method that emphasizes the scaling properties of the
underlying field equations.

In this paper we study the asymptotic evolution of vacuum $G_2$
metrics using a different approach than that used in the above
referred papers: we make use of renormalization group (RG) tools
to study the structurally stable characteristics of the
inhomogeneous metrics. Recently, RG techniques have been exhibited
as a powerful implement to study the asymptotic behavior of
partial differential equations~\cite{BKL,BK2,CGOI,CGOII}. The RG
method has been applied to study  homogeneous and isotropic
universe, an spherically symmetric dust collapse~\cite{IHK},
critical phenomena related with gravitational collapse~\cite{KHA},
Newtonian cosmology~\cite{SKM}, homogeneous flat causal bulk
viscous cosmological models~\cite{BHM} and the theory of
perturbations of isotropic universe with dynamically evolving
Newton constant and cosmological constant~\cite{RW,Sola}.

The plan of the paper is the following: in Section~\ref{rgmethod}
we illustrate the application of the RG method to a homogeneous
Bianchi type metric with a scalar field. This case has been
studied before and we recover the attractors of the system by
means of the RG method. In Section~\ref{Inhomogeneous} we apply
the RG technique to the vacuum, diagonal, $G_2$ metric. We find
the fixed-points and analyze their stability as well. We conclude
with Section~\ref{discuss}.

\section{The RG method: an illustrative example} \label{rgmethod}

It is well known that the asymptotics of partial differential
equations can often be found from the consideration of scaling
solutions (the equivalence of RG theory and the theory of
intermediate asymptotics was shown by Goldenfeld et.
al.~\cite{GMO1989}). Though it is usual practice to find the
similarity variable to analyze scaling solutions from a
combination of variables using dimensional arguments, however,
there is a large class of problems where this cannot be done
\cite{Barenblatt}. RG provides a systematic approach for finding
the scaling variables as well as the asymptotics of partial
differential equations.

The general (RG) method adapted to partial differential equations
is well known~\cite{BKL}. Therefore, instead of reproducing the
whole procedure, we explain this method using an illustrative
example which has been studied with other methods and has a
non-trivial asymptotic structure. We consider a class of
anisotropic cosmological models given by~\cite{CIH}
\begin{equation}\label{space-time}
ds^2 = -dt^2+a(t)^2 dx^2+ b(t)^2 e^{2mx} dy^2 +c(t)^2 e^{2x} dz^2
\; .
\end{equation}
This represents a one parameter ($m$) family of Bianchi models
with Bianchi $\text{III}, \text{V}$ and $\text{VI}_{0}$ for,
respectively, $m = 0, 1$ and $-1$, and Bianchi
$\text{VI}_{\text{h}}$ ($h=m+1$) for all other vlaues.

For a direct comparison with earlier works we choose the same
variables, namely, shear ($\sigma$) and expansion ($\theta$) for
our analysis. Note, however, that the feasibility of the earlier
analysis depends crucially on finding the ``right" set of
variables to work-with, otherwise system is too complex to analyze
completely, even in this homogeneous case. Moreover, the choice of
these variables is model dependent and rather ad-hoc. Therefore,
one also needs to be lucky to be able to analyze. Whereas, with RG
one can directly work with the metric functions, as  will be
done in the next Section.

The shear and expansion for this spacetime are given by
\begin{eqnarray}\label{sh-exp}
\theta & = & \frac{\dot a}{a}+\frac{\dot b}{b} +\frac{\dot c}{c}
\; ,\\
{\sigma}^2 & = & \frac{1}{3}\left[\left(\frac{\dot a}{a}\right)^2
+ \left(\frac{\dot b}{b}\right)^2 + \left(\frac{\dot
c}{c}\right)^2\right] .
\end{eqnarray}
Here an over-dot signifies derivative with respect to $t$. We
consider energy momentum tensor describing a minimally coupled
homogeneous scalar field $\phi(t)$, given by
\begin{equation}\label{em-tensor}
T_{\mu \nu} = \phi_{,\mu} \phi_{,\nu} - g_{\mu \nu}
\left(\frac{1}{2}\phi_{,\alpha}\phi^{,\alpha}+V(\phi)\right) ,
\end{equation}
with an exponential potential $V(\phi)=\lambda \exp(k\phi)$. Here,
$\lambda > 0 $ and $k$ are both constants. The evolution equations
are then given by
\begin{eqnarray}\label{e-eqns}
{\dot \theta} & = & -2\sigma^2 - \frac{\theta^2}{3} - {\dot
\phi}^2 +V(\phi) \; , \label{e-eqns1}\\
{\dot \sigma} & = & -\sigma \theta + p(m) \left( \theta ^2 -3
\sigma ^2 -\frac{3}{2} {\dot \phi}^2 -3 V(\phi) \right) \; ,
\label{e-eqns2} \\
{\ddot \phi} & = & - \theta {\dot \phi} - k V(\phi) \; ,
\label{e-eqns3}\\
{\dot V} & = & k {\dot \phi} V \; ,\label{e-eqns4}
\end{eqnarray}
where $p(m)=(1-m)/(3 \sqrt{3(1+m+m^2)})$. Since above system is an
autonomous system of differential equations, in the previous
study~\cite{CIH} the theory of dynamical systems was used to
analyze its asymptotic behavior using ``expansion-normalized
variables". We will use, as an alternative method, the RG
technique and recover the same results. Again, our goal is to
illustrate the method in this section and is applied it to a new
scenario in next section.

We find it useful to work with a compact notation and, therefore,
define a new indexed variable $u_i(t)$ that signifies the set
\{$\theta, \sigma, {\dot \phi}, V $\}, for $i=1,2,3,4$,
respectively. Our interest is in the asymptotics of solution of
the form
\begin{equation}\label{asymptot}
\lim_{t\rightarrow \infty} u_i(t) = t^{-\alpha_i} u_i^*(1) \; ,
\end{equation}
where argument ``$1$" signifies initial value of the quantity. It
is convenient to fix initial time as $t=1$. Now we will illustrate
how RG method gives a systematic procedure to fix $\alpha_i$ and
determine the scaling function $u_i^*(1)$ as fixed-points of RG
equations (in a inhomogeneous case the argument of $u_i^*$ will
not be a constant but will depend on a combination of time and
spatial variable, the scaling variable).

Let us consider transformations of the form
\begin{eqnarray}  \label{scaling}
t& \longrightarrow & L t \; ,  \notag \\
u_i(t)& \longrightarrow & U_i(t) = L^{\alpha_i} u_i(Lt) \; ,
\notag \\
\end{eqnarray}
which leave the equations invariant, i.e., if $u_i(t)$ is a
solution so is $U_i(t)$. We use a number $L>1$ as a parameter of
scale transformation. Note that unlike in the application of RG
method to quantum field theories or statistical mechanics, there
is no natural way to choose a scale $L$ here. Moreover, $L$ never
appears explicitly in the equations.

From equation set (\ref{e-eqns1})-(\ref{e-eqns4}) and
(\ref{scaling}) it is straightforward to recover
\begin{equation}\label{fix-const}
\alpha_1 = \alpha_2 = \alpha_3 = \frac{\alpha_4}{2} = 1 \; .
\end{equation}
Now, defining $L=\exp(\tau)$, and using Eq. (\ref{scaling}) in
(\ref{e-eqns1})-(\ref{e-eqns4}) we get a new set of evolution
equations:
\begin{equation}
\frac{dU_i}{d\tau}=\alpha_i U_i + \frac{\partial U_i}{\partial t}
\; .
\end{equation}
The procedure of defining $\tau$ is, in a sense, analogous to
summing over all degrees of freedom corresponding to fluctuations
of scale less than $L$ and then re-scaling everything by $L^{-1}$.
This new set of evolution equations define the RG transformations.
Since scaled quantities satisfy same evolution equations the time
derivative in the equation set above can be replaced with
equations (\ref{e-eqns1})-(\ref{e-eqns4}), and we have
\begin{eqnarray}\label{new-evolution}
\frac{d U_1}{d\tau} & = & U_1 -2 U_2^2
-\frac{{U_1}^2}{3}-{U_3}^2+{U_4} \; , \\
\frac{d U_2}{d\tau} & = & U_2 -U_1 U_2 +p(m)
\left({U_1}^2 -3{U_2}^2 -\frac{3}{2}{U_3}^2-3{U_4}\right) \; , \\
\frac{d{U_3}}{d\tau} & = & (1- U_1) {U_3} -{k U_4} \; , \\
\frac{d{U_4}}{d\tau} & = & (2+k {U_3}){U_4} \; .
\end{eqnarray}
All the quantities on the right hand side are evaluated at $t=1$.
Scale-invariant solutions emerge now from the fixed-point
structure of the RG map, which is defined by
\begin{equation}\label{fixed-point}
\frac{d{U_i}^*}{d\tau} = 0 \; ,
\end{equation}
with ${U_i}^*$ being the fixed-points. The complete set of
fixed-points for this system are given in the table below :
\vspace{0.3in}
$$
\begin{array}{|c|c|c|c|c|}
\hline  & U_1 & U_2 & U_3 & U_4 \\
\hline 1. & 1 & \pm \sqrt{\dfrac{2-3U_3}{6}} & U_3 & 0 ,\\
 \hline 2. & \dfrac{6}{k^2} & 0 & -\dfrac{2}{k} &
2\dfrac{(6-k^2)}{k^4} \\
\hline 3. & \dfrac{2(1-m)^2}{27 \text{p(m)}^2(1+m^2)} &
\dfrac{(1-m)^2}{9\text{p(m)}(1+m^2)}& 0 & 0 \\
\hline 4. & \left(1+\dfrac{k^2}{27 \text{p(m)}^2} \right)
\dfrac{2(1-m)^2}{k^2(1+m^2)} & \dfrac{(k^2-2)}{9k^2 \text{p(m)}}
\dfrac{(1-m)^2}{(1+m^2)}& -\dfrac{2}{k}&
\dfrac{2(1+m)^2}{k^2(1+m^2)}+\dfrac{4(1-m)^2}{k^4(1+m^2)} \\
\hline
\end{array}
$$
\centerline {Table I: Fixed-points of Bianchi $VI_{h}$ models}
\vspace{0.3in}

Therefore, we have four exact scale invariant solutions of the form
(\ref{asymptot}) with exponents $\alpha_i$ given by
(\ref{fix-const}) and $u_i^*(1)$ being the fixed-points given
above. Since the stability properties of these solutions is well
studied in literature~\cite{CIH}, we move now to the more general
inhomogeneous case. We analyze the fixed point structure as well
as their stability properties, which is not done before.

\section{Inhomogeneous Case}\label{Inhomogeneous}

We begin with a family of metrics which have been a very useful
tool for studying inhomogeneous cosmological models, known as the
generalized Einstein-Rosen spacetimes (see for example,
\cite{verdaguer, CCM}). The line element is of the form:
\begin{equation}
ds^{2} = f^2(dz^2 - dt^2) + \gamma_{ab} dx^a dx^b, \; \; a,b = 1,2 \; .
\label{metric}
\end{equation}
Here $x^1=x$, $x^2=y$, and both $f$ and $\gamma_{ab}$ are
functions only of $z$ and $t$. Metric (\ref{metric}) is fairly
general and includes Bianchi type models I to VII. In this paper we are
interested in vacuum solutions only. We can impose
now
\begin{equation}  \label{condition}
\mbox{det}\gamma_{ab}=t^2 \;,
\end{equation}
since in vacuum case one of the field equation takes the form
\begin{equation}  \label{justify}
(\mbox{det}\gamma_{ab})^{(1/2)}_{,tt}-(\mbox{det}
\gamma_{ab})^{(1/2)}_{,zz}=0 \; .
\end{equation}
$t$ and $z$ are two independent solutions of this equation. Under
this condition the coordinates $t$ and $z$ are called as the
canonical coordinates; and there is no loss of generality in this
choice~\cite{verdaguer}. To simplify the analysis we now
specialize to the diagonal metrics, i.e., metrics with a single
polarization.

The spacetime can be written in the form
\begin{equation}  \label{sptm}
ds^2=f^2(dz^2-dt^2)+t(h^2 dx^2 + h^{-2} dy^2) \; ,
\end{equation}
where $f$ and $h$ are functions of $t$ and $z$ only. These metrics
admit an Abelian $G_2$ group of isometries with 2
spacelike commuting killing vectors $\partial_x$ and $\partial_y$.
The set of Einstein field equations in the vacuum case reduces to
\begin{eqnarray}
h_{,t}&=&\frac{1}{2t} \frac{f_{,z}}{f} \frac{h^2}{h_{,z}},  \label{dht} \\
f_{,t}&=&-\frac{1}{4t}f + t f \left(\frac{h_{,z}}{h}\right)^2 +
\frac{1}{4t} \frac{(f_{,z})^2}{f} \left(\frac{h}{h_{,z}}\right)^2
.  \label{dft}
\end{eqnarray}
Eqs. (\ref{dht}) and (\ref{dft}) are the evolution equations for
$h$ and $f$ , respectively, and it is easy to check that the
remaining field equations are identically satisfied. We now follow
the prescription given in the previous section to recover the
exact scaling solutions.

Let us consider the following scale transformation
\begin{eqnarray}  \label{Ltrans}
z& \longrightarrow & L z \; ,  \notag \\
t& \longrightarrow & L^{\epsilon}t \; ,  \notag \\
h(t,z)& \longrightarrow & \phi(t,z)=L^{\alpha}
h(L^{\epsilon}t,Lz) \; ,  \notag \\
f(t,z)& \longrightarrow & \psi(t,z)=L^{\beta} f(L^{\epsilon}t,L z)
\; .
\end{eqnarray}
Here $\phi$ and $\psi$ are the scaled quantities. Since scaled
quantities also satisfy the original equations, Eq. (\ref{dft})
fixes
\begin{equation}  \label{vepsilon}
\epsilon = 1 \; .
\end{equation}
The scaling relations (\ref{Ltrans}) along with (\ref{vepsilon})
and successive transformations, first $t\rightarrow 1$ and then $L
\rightarrow t$, gives scale invariant solution of the form
\begin{eqnarray}  \label{nsol}
h(t,z)&=&t^{-\alpha}\phi(1,z/t) \; ,  \notag \\
f(t,z)&=&t^{-\beta}\psi(1,z/t) \; .
\end{eqnarray}
The equations above express arbitrary solution in terms of initial
data (at $ t=1$). As stated earlier we work with $t=1$ as our
initial time and evolution is in sense of scaled time $Lt$ with
$L>1$.

{}From simple dimensional analysis we would have $\beta=0$ and
$\alpha = \pm 1/2$ (introducing a dimensional constant multiplying
either $dx$ or $dy$). However, as we will see later this would
lead to a trivial solution: the homogeneous Kasner metric. The
fact that we recover Kasner solution as a fixed point of a general
inhomogeneous G2 metric though is non-trivial and it elucidates
the appearance of this solution in earlier studies; it is used as
a seed metric in different solution generating techniques, and
more importantly, is known to describe ``generic" cosmological
singularity in the analysis of Belinskii et al. \cite{LBK}. In
order to get a non-trivial structure in fix-point analysis we need
to analyze the anomalous dimensions for both the functions $f$ and
$h$. It is well known \cite{NGoldenfeld} that the anomalous
dimensions $\alpha \neq \pm 1/2$ and $\beta \neq 0$ are fixed by
initial or boundary conditions. In our case, cosmological vacuum
solutions, initial condition refers to geometry at $t=0$. Since,
metrics of the type (\ref{sptm}) are singular at $t=0$ there is
not strict functional constraint on the system described by
(\ref{dht}) and (\ref{dft}). Nevertheless, interpreting a
posteriori the fixed points we can give some hints to understand
the role of anomalous dimensions $\alpha$ and $\beta$.

Denoting $L=\mbox{exp}(\tau )$ the RG equations are
\begin{eqnarray}
\frac{d \phi}{d\tau }&=&\alpha \phi+{\phi^{\prime }}
z+\frac{1}{2}\left( \frac{{\psi^{\prime }}}{ \psi}\right) \left(
\frac{\phi^{2}}{{\phi^{\prime }}}\right) \; ,   \label{RGeq1} \\
\frac{d{\psi}}{d\tau } &=&\left( \beta -\frac{1}{4}\right) \psi+
\psi^{\prime }z+\psi\left( \frac{{\phi^{\prime
}}}{\phi}\right)^{2}+\frac{1}{4}\frac{({\psi^{\prime
}})^{2}}{\psi} \left(\frac{{\phi}}{{\phi^{\prime }}}\right) ^{2}
\; .
\end{eqnarray}
We now investigate the fix-point structure of the equation set
above, i.e.,
\begin{eqnarray}
\frac{d{\phi^{\ast }}}{d\tau }=0 &\Rightarrow &\alpha =-\left(
\frac{\phi^{\ast '}}{{\phi^{\ast }}} \right) z-\frac{1}{2}\left(
\frac{\psi^{\ast '}}{{ \psi^{\ast }}}\frac{{\phi^{\ast
}}}{\phi^{\ast '}}\right) \; ,  \notag  \label{fixpt} \\
\frac{d{\psi^{\ast }}}{d\tau }=0 &\Rightarrow &\frac{1}{4}-\beta
=\left( \frac{\psi^{\ast '}}{{\psi^{\ast }}}\right) z+\left(
\frac{\phi^{\ast '}}{\phi^{\ast }}\right)^{2}+\frac{1}{2}\left(
\frac{\psi^{\ast '}}{{\psi^{\ast }}}\frac{\phi^{\ast }}{\phi^{\ast
'}}\right) ^{2} \; .
\end{eqnarray}
System decouples to give
\begin{eqnarray}
\left( \frac{{\phi^{\ast '}}}{{\phi^{\ast } }}\right) & = & \pm
\sqrt{\frac{\triangle}{{z^{2}-1}}} \; ,
\label{decoupleh} \\
\left(\frac{\psi^{\ast '}}{{\psi^{\ast } }}\right) & = &
\frac{2\sqrt{\triangle}}{1-z^2} \left(\sqrt{\triangle} z\pm
\alpha\sqrt{z^2-1}\right) \;, \label{decouplef}
\end{eqnarray}
where $\triangle =\alpha ^{2}+\beta -1/4$. The real solutions
correspond to $\triangle > 0$ for $z^2 > 1$ and $\triangle < 0$
for $z^2 < 1$. The equations (\ref{decouplef}) and
(\ref{decouplef}) can be easily integrated to give the
fixed-points
\begin{eqnarray}
{\phi^{\ast }} & = &(z+\sqrt{z^{2}-1})^{\pm\sqrt{\triangle} } \; ,
\label{crit} \\
{\psi^{\ast }} & = & c_f (z+\sqrt{z^{2}-1})^{\mp 2\alpha\sqrt{
\triangle} } \; (z^{2}-1)^{-\triangle } \; ,
\end{eqnarray}
were $c_f$ is an integration constant. We have dropped constant of
integration from ${\psi^{\ast }}$ since this can be absorbed
simply by scaling of $x$ and $y$. Also, we note here that the new
parameter $\triangle $ relates our spacetime metric with Kasner
metric in the following way. The Kasner metric can be written in
the form
\begin{equation}
ds^{2}=t^{(d^{2}-1)/2}(dz^{2}-dt^{2})+t^{1+d}dx^{2}+t^{1-d}dy^{2},
\label{kasner}
\end{equation}
where parameter $d$ can be chosen positive or negative. A direct
comparison with the metric (\ref{sptm}) using (\ref{nsol}) gives
the Kasner relationship
\begin{equation}
4\alpha ^{2}+4\beta -1=0 \; .  \label{krel}
\end{equation}
Therefore, $\triangle=0$ case corresponds to the Kasner models.
When $\triangle > 0$ ($z^2 > t^2$) the critical solution is:
\begin{eqnarray}
h(t,z) &=&t^{-\alpha \mp \sqrt{\triangle}
}(z+\sqrt{z^{2}-t^{2}})^{\sqrt{\triangle} } \;, \notag
\label{crit-sol} \\
f(t,z) &=&c_f
t^{(\alpha\pm\sqrt{\triangle}+1/2)(\alpha\pm\sqrt{\triangle}-1/2)}
\; (z+\sqrt{z^{2}-t^{2}})^{\mp 2\alpha \sqrt{\triangle} } \;
(z^{2}-t^{2})^{-\triangle } \; .
\end{eqnarray}
and when $\triangle < 0$ ($z^2 < t^2$) the critical solution is:
\begin{eqnarray}
h(t,z) &=& t^{-\alpha}\exp(\mp\sqrt{-\triangle}\arccos\frac{z}{t})
\; , \notag \label{crit-sol2} \\
f(t,z) &=& c'_f t^{\alpha^2+\triangle-1/4} (t^2-z^2)^{-\triangle}
\exp(\pm 2\alpha\sqrt{-\triangle}\arccos\frac{z}{t}) \;.
\end{eqnarray}

The spacetime is split into two regions separated by the
light-cone $z = t$. In each region the solution takes one of the
forms given by the expressions above. For $\triangle =0$ we have
the Kasner metric. In the general case ($\triangle \neq 0$),
solution (\ref{crit-sol}) is the soliton metric that has been
obtained earlier by the inverse scattering transformation with
real poles from the Kasner metric, and the solution
(\ref{crit-sol2}) is the cosoliton solution generated also from
the Kasner metric. Both solutions have been studied in~\cite{V-B}
(and references therein). Depending of the parameters, the
light-cone $z = t$ is singular for the solution (\ref{crit-sol})
but is always singular for (\ref{crit-sol2}). This means that even
though the metric and its first derivatives are continuous across
the light-cone, the solutions cannot be matched across the
light-cone (for a discussion on the matching of these metrics see
\cite{V-B}). We would like to note here that the $\pm$ sign in the
above equations is actually related to the symmetry under $
x\leftrightarrow y $, and as we will see later in the stability
analysis same results hold for both the signs. Moreover, we would
like to stress here that the solutions which we have obtained are
actually the future asymptotic states, i.e., to which spacetime
``prefers" to settle down. This makes this analysis very powerful
since not only we recover in a very simple fashion a whole class
of scale invariant solutions but also, due to ``universality",
these solutions actually are the preferred asymptotic states.

We will consider now the linear stability analysis. Since these
solutions emerge as fixed-points of the RG map, the stability of
these solutions is the stability of these fix-points. Let us
define
\begin{eqnarray}\label{def-pert}
\phi &=& \phi^*(1+\delta\phi) \; , \\
\psi &=& \psi^*(1+\delta\psi) \; ,
\end{eqnarray}
where $\delta\phi \ll 1$  and $\delta\psi \ll 1$. The above form
of perturbations is chosen to facilitate the analysis since both
unperturbed functions $\phi^*$ and $\psi^*$ can diverge at $z= \pm
1$ and $z=\pm \infty$.

The perturbation equations takes the form
\begin{eqnarray}\label{pert-equations}
\frac{d\delta\phi}{d\tau} &=&
\left(\frac{\phi^*}{{\phi^*}'}\right)\left[\left(\alpha
+2z\frac{{\phi^*}'}{\phi^*} \right) \frac{d\delta\phi}{d z} +
\frac{1}{2}\frac{d\delta\psi}{d z}\right] \; , \notag \\
\frac{d\delta\psi}{d\tau} &=&-
\left(\frac{\phi^*}{{\phi^*}'}\right)\left[2\left(\alpha^2+\triangle
+2\alpha z\frac{{\phi^*}'}{\phi^*} \right) \frac{d\delta\phi}{d z}
+ \alpha \frac{d\delta\psi}{d z} \right] \; .
\end{eqnarray}
We shall compute normal modes assuming the following form for the
perturbations:
\begin{equation}\label{n-m}
\delta\phi=e^{\omega\tau}\rho(z) \; ,\qquad
\delta\psi=e^{\omega\tau}\sigma(z) \; ,
\end{equation}
where $\omega$ is a constant. The perturbation equations
(\ref{pert-equations}) can be written as:
\begin{eqnarray}\label{pert-equations2}
\left(\alpha
+2z\frac{{\phi^*}'}{\phi^*} \right)\rho' +
\frac{1}{2}\sigma' & = & \frac{{\phi^*}'}{\phi^*}\omega\rho \; ,\notag \\
-2\left(\alpha^2+\triangle +2\alpha z\frac{{{\phi^*}'}}{\phi^*}
\right)\rho' - \alpha \sigma' & = &
\frac{{\phi^*}'}{\phi^*}\omega\sigma \; .
\end{eqnarray}
From these equations it is easy to see that:
\begin{equation}\label{sigma}
\omega\sigma=-2\sqrt{\triangle}\sqrt{z^2-1} \rho'-2\alpha \omega
\rho \; .
\end{equation}
Taking the derivative of this equation and substituting in
(\ref{pert-equations2})  we get:
\begin{equation}\label{rho}
\rho''+\frac{z}{z^2-1}(1-2\omega)\rho'+\frac{\omega^2}{z^2-1}\rho=0
\; .
\end{equation}
Thus, the problem of solving linear perturbations around the
fixed-points has been reduced to finding solutions of the above
equation, and (\ref{sigma}) completes the solution. The general
solution of (\ref{rho}) is given by:
\begin{equation}\label{legendre}
\rho(z)=(z^2-1)^{\omega/2+1/4}\left(c_1P_{-1/2}^{\omega+1/2}(z) +
c_2Q_{-1/2}^{\omega+1/2}(z)\right) \; ,
\end{equation}
where $P^\mu_\nu$ and $Q^\mu_\nu$ are Legendre functions of first
and second kind, respectively, and $c_1$ and $c_2$ are arbitrary
constants. From (\ref{sigma}) the solution for $\sigma$ can be
easily obtained:
\begin{eqnarray}\label{legendre2}
\sigma(z)=& & -2\sqrt{\triangle}(z^2-1)^{\omega/2-1/4}
\left[c_1\left(z P_{-1/2}^{\omega+1/2}(z)-
P_{1/2}^{\omega+1/2}(z)\right) + c_2
\left(zQ_{-1/2}^{\omega+1/2}(z)-
Q_{1/2}^{\omega+1/2}(z)\right) \right]\notag\\
 &  & -2\alpha(z^2-1)^{\omega/2+1/4} \left(c_1P_{-1/2}^{\omega+1/2}(z)
 + c_2Q_{-1/2}^{\omega+1/2}(z)\right) \; .
\end{eqnarray}
We require that the perturbations be regular for all $z$. Since
the differential equation (\ref{rho}) has four regular singular
points at $z=\pm 1, \pm \infty$ we must consider three different
regions: $1<z<\infty$, $-\infty<z<-1$ and $-1<z<1$. The first and
the second region are equivalent. Let's first analyze the behavior
in the first region. It is easy to see that the leading terms of
the two linearly independent solutions $\rho_1$ and $\rho_2$ of
(\ref{rho}) in the neighborhood of $z=\infty$ are respectively
$\rho_1\approx z^\omega + O(z^{\omega-2})$ and $\rho_2\approx
\rho_1 \ln z + O(z^{\omega-2})$. Both solutions are regular if
$\omega\leq0$. Furthermore, from (\ref{sigma}), $\sigma$ is also
regular provided that $\omega\leq 0$. At $z=1$ the leading terms
of the two linearly independent solutions for $\rho$ are:
$(z-1)^{\omega+1/2}+O((z-1)^{\omega+3/2})$ and
$constant+O((z-1))$. For $\sigma$ the leading terms are:
$(z-1)^\omega$ and $constant+O((z-1))$.  Since $\omega$ is
negative the first independent solution must be neglected so that
we have regular solutions. For instance, tacking $c_2=0$ in
(\ref{legendre}) the solution is regular.

Making the scale transformation given by
(\ref{Ltrans})-(\ref{nsol}) we get the the behavior of the
solution near the fixed-point :
\begin{eqnarray}
\delta \phi & = & t^\omega \rho(z/t) \; ,\notag\\
\delta \psi & = & t^\omega\sigma(z/t) \; ,
\end{eqnarray}
with $\omega\leq 0$ and $ z>t$.  Since the solution above is valid
in the region $z>t$, taking limit $t\rightarrow\infty$ means
$z\rightarrow\infty$ as well. We can therefore distinguish two
different situations. First, when we approach infinity in such a
way that we encounter the light-cone $z=t$ without crossing it. In
this case asymptotically $z/t=1$ and, bearing in mind the behavior
of the regular solutions of (\ref{rho}) close to the point $z=1$,
the above perturbations behave as:
\begin{equation}
\delta\phi\approx\delta\psi\approx constant\times t^\omega \; .
\end{equation}

In the second situation we approach infinity through a path that
does not encounter the light-cone. In this case $z/t$ tends to a
constant greater than 1, and the perturbations behave in a similar
way that that given by the above expressions. So, in both cases
the perturbations tend to zero when $t\rightarrow \infty$ which
means that the fixed solution outside the light-cone is stable.

Let's perform this analysis in the region $-1<z<1$. The leading
terms of the independent solutions of (\ref{rho})close to $z=1$
are those described earlier for the same point changing $z-1$ by
$1-z$. When $z=-1$ the behavior is the same changing $z-1$ by
$z+1$. From this we can not obtain a definite conclusion about the
values of $\omega$ which make the solution regular. To do that is
convenient to transform the equation (\ref{rho}) to a
hypergeometric differential equation with coefficients $a=b=1/2$
and $c=1/2-\omega$. It is not difficult to see that one of the
independent solutions the solution of (\ref{rho}) and
(\ref{sigma}) is given by:
\begin{eqnarray}
\rho(z) & = & c_1(1+z)^{\omega+1/2} \;F\left(\frac{1}{2},
\frac{1}{2};\frac{1}{2}-\omega;\frac{1-z}{2}\right) \; ,\notag\\
\sigma(z) & = &
-2c_1\frac{\sqrt{|\triangle|}}{\omega}\sqrt{1-z^2}(1+z)^{\omega-1/2}
\left[
(\omega+\frac{1}{2}) F\left( \frac{1}{2}, \frac{1}{2};
\frac{1}{2}-\omega; \frac{1-z}{2}\right)-\frac{1}{4-8\omega} (1+z)
F\left( \frac{3}{2}, \frac{3}{2}; \frac{3}{2}-\omega;
\frac{1-z}{2}\right)\right]
\notag\\
 & & -2\alpha c_1(1+z)^{\omega+1/2} \;F\left(\frac{1}{2},
\frac{1}{2};\frac{1}{2}-\omega;\frac{1-z}{2}\right) \; .
\end{eqnarray}
The form of the second independent solution depends on the values
of $\omega$. In any case, the above solution is regular when
$\omega$ is positive, except when $\omega=n+1/2$ ($n$ a positive
integer). Therefore,
with $ |z|<t$, the modes with $\omega$ positive
dominate and the solution in the interior of the light-cone is not
stable.

\section{Discussion} \label{discuss}

In this paper we have shown how the RG method can be used to
obtain the asymptotic regime of cosmological solutions.  We have
first illustrated  the method applying it to homogenous
cosmological models with a scalar field. These models have been
extensively analyzed using, mainly, the so called ``expansion
normalized variables"; which are a set of dimensionless
variables~\cite{W}. In those works the role of self-similar
solutions in describing the asymptotic regime has been stressed.
Since a fixed-point of the RG transformation is a scale-invariant
solution, it is reasonable to think that the RG method will
naturally render the same results (and we have manifestly shown
this here, section~\ref{rgmethod}) as those obtained using the
``expansion normalized variables". It is worthy to stress also
that those normalized variables are well adapted to the
dimensional and similarity analysis~\cite{Barenblatt}. These
results bring the asymptotic behavior of the homogeneous
cosmological models in a new perspective. Moreover, its simplicity
as well as  systematic approach show that it is a practical idea
to implement RG for studying asymptotic behavior of spacetimes in
General Relativity.

We have also applied the RG method to diagonal, vacuum
inhomogeneous $G_2$ metric. In this case, the system reduces to a
set of coupled partial differential equations whose analysis using
the RG method is not difficult. The fixed-point is an exact
solution depending on two parameters.  This solution belongs to
the class of soliton solutions. Soliton solutions are intended as
those solutions which can be obtained by the inverse scattering
transformation from  a known one~\cite{verdaguer}. In particular,
the found fixed-point is a soliton solution with real poles with
origin at $z=0$, whose ``seed" metric is the Kasner metric
(soliton origin marks the origin of the light-cone $z^2 = t^2$).
Since the metric (\ref{sptm}) is invariant under a $z$ translation
the fact that the origin is at $z=0$ is not important. Moreover, a
more general class of soliton solutions is that corresponding to a
sum of solitons each with a different origin. Since the RG method
gives the long-time behavior of the solution, the difference in
the origins tend to zero as $t$ tends to infinity leaving only one
origin. It is interesting to note that, generically, the solution
doesn't homogenize, let alone isotropize, for the final state is
inhomogeneous. There is, however, a few particular cases for which
the metric tends to a homogeneous solution: when $\triangle=0$ the
solution is the Bianchi type I Kasner metric, and for a particular
value of the parameters $\alpha$ and $\triangle$ the solution is
Ellis and McCallum family of vacuum Bianchi models~\cite{EM}.
Finally, we would like to note that the fixed-points recovered in
the RG technique gives all the scale-invariant exact solutions of
vacuum $G_2$ cosmologies.

Now we can understand the relation of the anomalous dimensions
$\alpha$ and $\beta$ with initial conditions: exponent $\alpha$
gives the Kasner parameter of the ``seed" metric to generate the
soliton solution by means of inverse scattering technique. The
parameter $\triangle$ (which fixes $\beta$) determines the number
of solitons in our solution. So, the two ``initial conditions"
used in the inverse scattering technique, number of solitons and
the Kasner seed metric, fixes the anomalous dimension of system
comprising of equations (\ref{dht}) and (\ref{dft}).

The RG method allows us to study the linear perturbations around
the fixed-points as well. We have shown that the solution out-side
the light-cone, which corresponds to the soliton solution with
real poles, is stable against bounded perturbations, contrary to
the solution inside the light-cone (cosoliton solution) which
we find to be unstable. Let's note that although the stability
analysis doesn't have any apparent dependence on the $\triangle$
parameter one should be very cautious to extend these results to
the case $\triangle=0$. This case corresponds to the Kasner
solution and this is the only case for which the fixed-point
results in ${\phi^*}'={\psi^*}'= 0$, that makes the system of
equations for the perturbations singular.

There remain a few open questions that we hope to deal-with in
future works. First is related with the role of soliton solutions
with complex poles. These can be obtained as a complex translation
along the $z$ axis of the soliton solution with real poles and
they are regular in all the spacetime, even on the light-cones.
Since this complexification doesn't alter the structure of the
field equations it would be worthy to investigate whether the
complex pole soliton solutions can represent an asymptotic state
of the inhomogeneous metrics. Second, it would be interesting to
extend the analysis performed in this paper to general
inhomogeneous non-vacuum metrics. Solutions with a scalar field
are of particular interest in order to investigate issues like
isotropization, scaling solutions, etc.. Finally, this technique
can be used to analyze long time behavior of spacetimes with more
that four dimensions, for example, brane-world scenarios.

\section*{acknowledgements}
We thank Takahiro Tanaka, Oriol Pujolas, M. Valle and J. A. Oteo
for discussions. J.I. acknowledges financial support under CICYT
grant BFM 2000-0018.

\end{document}